\documentclass[aps,twocolumn,prl,showpacs]{revtex4}
\usepackage[dvips]{graphicx}
\usepackage{latexsym}
\usepackage{epsfig}
\usepackage{bm}
\usepackage{amsmath,times,psfrag,subfigure}
\newcommand{\uvec}[1]{\hat{\mathbf{#1}}} 
\newcommand{\coloneq}{\; \colon \mspace{-12.0mu} =}

\begin{document}

\title{Flexoelectric blue phases}

\author{G. P. Alexander and J. M. Yeomans} 
\affiliation{Rudolf Peierls Centre for Theoretical Physics, 
University of Oxford, 1 Keble Road, Oxford, OX1 3NP, England.}  
\date{\today}

\begin{abstract}
We describe the occurence and properties of liquid crystal phases showing two dimensional splay and bend distortions which are stabilised by flexoelectric interactions. These phases are characterised by regions of locally double splayed order separated by topological defects and are thus highly analogous to the blue phases of cholesteric liquid crystals. We present a mean field analysis based upon the Landau--de Gennes ${\bf Q}$-tensor theory and construct a phase diagram for flexoelectric structures using analytic and numerical results. We stress the similarities and discrepancies between the cholesteric and flexoelectric cases.
\end{abstract}


\maketitle

Elastic distortions in the nematic phase of liquid crystals can be
categorised into three types: splay, twist and bend. Stable phases
with two dimensional twist distortions have been well characterised and
observed experimentally. Here we predict the occurence of structures
with two dimensional splay--bend distortions.

\begin{figure}[tb]
\begin{center}
\includegraphics[width=38mm]{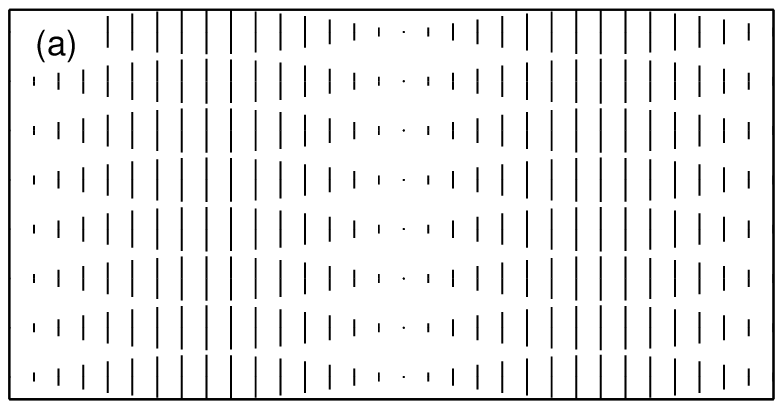} \hspace{4mm}
\includegraphics[width=38mm]{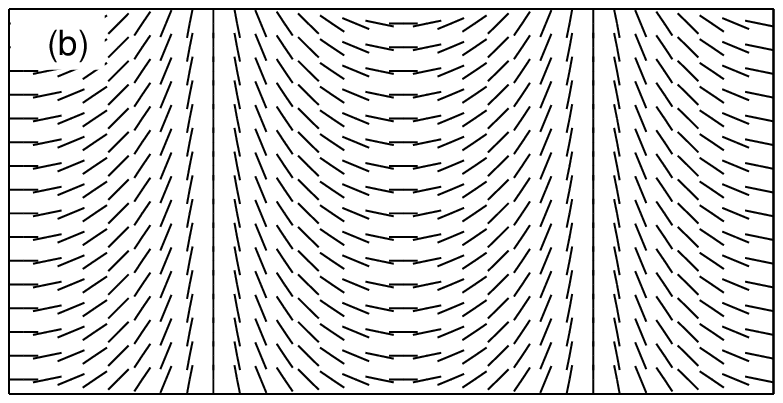}
\end{center}
\caption{Schematic representation of elastic distortions of nematic liquid crystals, showing the director field projected into the plane of the page. (a) Twist in
the cholesteric phase of chiral nematics. (b) Splay and bend in a
nematic with flexoelectric interactions.}
\label{fig:distortions}
\end{figure}

\begin{figure}[tb]
\begin{center}
\includegraphics[width=35mm]{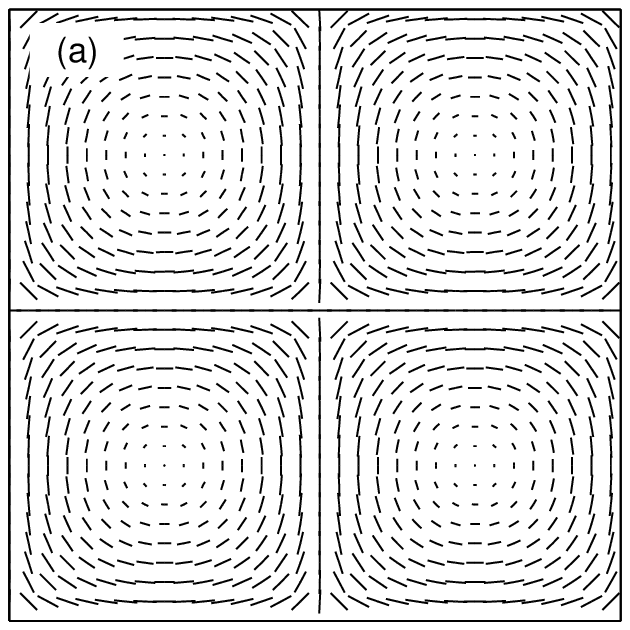} \hspace{5mm}
\includegraphics[width=35mm]{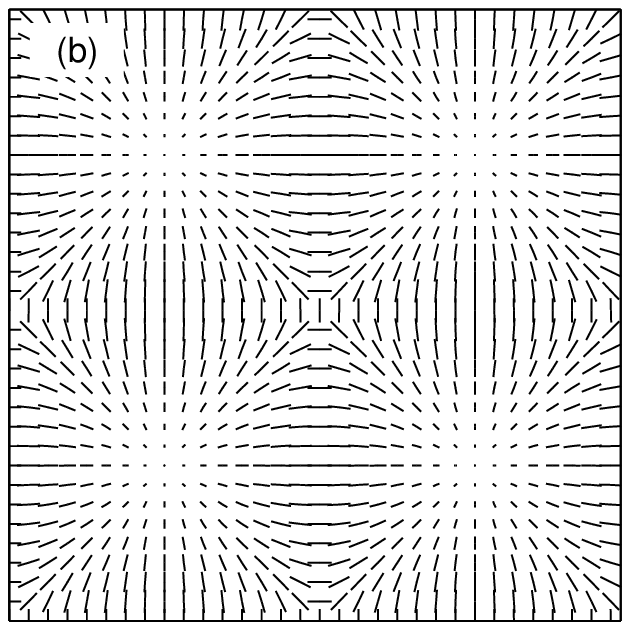}
\caption{Schematic examples of two dimensional director field configurations. (a) Structure displaying regions of local double twist, characteristic of cholesteric blue phases. (b) Analogous structure possessing double splay distortions.}
\label{fig:square}
\end{center}
\end{figure}

When nematics are doped with chiral molecules they can show stable phases with a natural twist known as cholesterics, shown in Fig.~\ref{fig:distortions}(a).
In general cholesteric phases comprise a {\em one}-dimensional helix. This is because topological constraints mean that it is not possible to construct a state with helical ordering in two dimensions without introducing defects, or disclination lines, into the structure. However local regions of double twist are possible. Fig.~\ref{fig:square}(a) shows a simple, two dimensional example of a director field where regions of double twist are separated by topological defects.
If the free energy advantage of the double twist regions offsets the
disadvantage of the disclinations phases of this type will
be stable. 

Such states, the so-called blue phases of cholesteric liquid crystals,
have indeed been observed \cite{wright}. The disclination structure is, however, more complicated than that of the simple example in Fig.~\ref{fig:square}(a). The disclinations form textures with cubic symmetry and the stable phases have space groups $O^{8-}$ and $O^2$.
The blue phases are stable at the isotropic--cholesteric phase
boundary as here the magnitude of the order is small and hence the free energy penalty associated with disclinations is low. They are stable only over a narrow temperature range $\sim \!1$ K, although recently this has been extended to $\sim \!60$ K by adding polymers or bimesogenic molecules \cite{kikuchi,coles}.

One might therefore ask whether analogous behaviour could be observed
in a liquid crystal which shows splay and bend, rather than twist,
distortions. In 1969 Meyer showed that the one dimensional splay--bend
distortion of the director field, shown in
Fig. \ref{fig:distortions}(b), could result from flexoelectric
coupling to an external field \cite{meyer}. Here we extend these
results to show that near the isotropic--nematic transition two
dimensional splay--bend structures (e.g., Fig.~\ref{fig:square}(b)) can be stable. We shall term such
phases flexoelectric blue phases. Similar director field configurations, but stabilised by a saddle splay term in the free energy, have been reported in \cite{chakrabarti}.

The coupling of a liquid crystal to an external field occurs in two principal forms. The most usual is dielectric, where the field couples directly to the orientational order parameter. This coupling is quadratic in the field strength and the molecules tend to align parallel or perpendicular to the field, depending on the sign of the dielectric anisotropy. Flexoelectricity, which we shall deal with here, occurs because of the elastic properties of liquid crystals. If the liquid crystal is composed of pear-shaped molecules an elastic distortion can lead to a spontaneous polarisation. Conversely, if a polarisation is induced by an external electric field then an elastic distortion results. This is termed flexoelectric coupling, and it gives a response linear in the field strength. 

Our conclusions are based upon Landau--de Gennes mean-field theory.
We present analytic and numerical results exploring the phase 
diagram of the flexoelectric blue phases, stressing the analogy, and differences, 
between these and the cholesteric blue phases. 

At a macroscopic level, the order parameter ${\bf Q}$ may be taken to be a traceless, symmetric, second rank tensor which is related to the anisotropic part of the dielectric tensor \cite{degennes}. Using a tensor allows both the magnitude and the direction of the order to be recorded. The direction of the order, described by a vector ${\bf n}$ called the director, is defined as the eigenvector corresponding to the maximal eigenvalue of ${\bf Q}$. We consider a bulk sample with periodic boundary conditions and assume that the liquid crystal has zero dielectric anisotropy and the electric field is uniform throughout the sample. 

The equilibrium thermodynamics of a nematic liquid crystal can be described by the Landau--de Gennes free energy \cite{degennes}, which we supplement by a familiar term for chiral coupling and an additional flexoelectric coupling
\begin{equation}
\begin{split}
F = \tfrac{1}{V} \int_{\Omega} d^3r \biggl\{ & \tfrac{\tau}{4} \, \text{tr}\bigl( \mathbf{Q}^2 \bigr) - \sqrt{6} \text{tr} \bigl( \mathbf{Q}^3 \bigr) + \Bigl( \text{tr}\bigl( \mathbf{Q}^2 \bigr) \Bigr)^2 \\
& + \tfrac{L_1}{2} \bigl( \nabla_a Q_{bc} \bigr)^2 + 2q_0L_1 {\bf Q}\cdot \nabla \times {\bf Q} \\
& + \varepsilon_f Q_{ab} \bigl( E_a \nabla_c - E_c \nabla_a \bigr) Q_{bc} \biggr\} \; .
\label{eq:freeenergy}
\end{split}
\end{equation}
Here, $\tau$ is a reduced temperature, $L_1$ is an elastic constant,
$q_0$ determines the pitch in the cholesteric phase,
$\varepsilon_f$ is a flexoelectric coupling constant, ${\bf E}$ is the
electric field and $V$ is the volume of the domain,
$\Omega$. There are four flexoelectric couplings up to second order in ${\bf Q}$, three of which may be written as a total divergence and converted to a surface term using Stokes' theorem. Since we neglect surface terms in this work only the term quoted need be retained. For simplicity, we adopt the one elastic constant approximation where the magnitude of twist, splay and bend are taken to be equal.

Since chirality and flexoelectricity induce structure in the fluid it is useful to analyse potential stable states by introducing a Fourier decomposition of the $\mathbf{Q}$-tensor \cite{brazovskii,ghsa}
\begin{equation}
\mathbf{Q}(\mathbf{r})\!=\! \sum_{\mathbf{k}} N_k^{-1/2}\! \sum_{m=-2}^2\! Q_m(k) \text{e}^{i\psi_m(\mathbf{k})} \mathbf{M}_m(\uvec{k}) \text{e}^{-i\mathbf{k}.\mathbf{r}} \; ,
\end{equation}
where $N_k$ is a normalising factor counting the number of wavevectors with the same magnitude, $Q_m(k), \psi_m(\mathbf{k})$ are the amplitude and phase of the Fourier component and ${\bf M}_m(\uvec{k})$ are a set of orthonormal basis tensors. Reality of $\mathbf{Q}$ is ensured by requiring $\psi_m(-\mathbf{k}) = - \psi_m(\mathbf{k})$ and $\mathbf{M}_m(-\uvec{k}) = \mathbf{M}^{\dagger}_m(\uvec{k})$. With these definitions the quadratic part of the free energy takes the form
\begin{equation}
\begin{split}
F^{(2)} & = \sum_{\mathbf{k}}  N_k^{-1} \sum_{m,m^{\prime}} Q_m(k)Q_{m^{\prime}}(k)\text{e}^{i\left( \psi_m(\mathbf{k}) - \psi_{m^{\prime}}(\mathbf{k}) \right)} \\
& \mspace{-20mu} \times M^{\dagger}_{m^{\prime}} \bigl( \uvec{k} \bigr)_{ab} \biggl\{ \Bigl( \tfrac{\tau}{4} + \tfrac{L_1}{2}k^2 \Bigr) \delta_{ac} - i 2q_0L_1 k \epsilon_{adc}\hat{k}_d \\
& \qquad \quad - i \varepsilon_f Ek \bigl( \hat{E}_a \hat{k}_c - \hat{k}_a \hat{E}_c \bigr) \biggr\} M_m \bigl( \uvec{k} \bigr)_{bc} \; .
\end{split}
\end{equation}
The basis tensors ${\bf M}_m(\uvec{k})$ should be chosen so as to
diagonalise $F^{(2)}$. It is convenient to introduce the
quantity $\Delta_{\mathbf{k}} \coloneq \smash[b]{\sqrt{1 - \bigl
( \uvec{E} . \uvec{k} \bigr)^2}}$ and use the direction of the
electric field and of the wavevector for a given Fourier mode to
define a local, right-handed, orthonormal frame field $\{ \uvec{k},
\uvec{v}, \uvec{w} \}$ such that $\uvec{E} = \bigl( \uvec{E}
. \uvec{k} \bigr) \uvec{k} + \Delta_{\mathbf{k}} \uvec{w}$. We then
make use of the observation that the flexoelectric coupling is
antisymmetric to rewrite it as $+ i \varepsilon_f E \Delta_{{\bf k}} k
\epsilon_{adc} \hat{v}_d$, revealing a formal mathematical similarity
between flexoelectric and chiral couplings. Combining these two
coupling terms motivates a transformation to a new set of
basis vectors, obtained by means of a rotation about $\uvec{w}$:
\begin{equation}
\begin{split}
\mathbf{a} & \coloneq \tfrac{1}{\mu_{\mathbf{k}}} \bigl( 2q_0L_1 \uvec{k} - \varepsilon_f E \Delta_{\mathbf{k}} \uvec{v} \bigr) \; , \\
\mathbf{b} & \coloneq \tfrac{1}{\mu_{\mathbf{k}}} \bigl( \varepsilon_f E \Delta_{\mathbf{k}} \uvec{k} + 2q_0L_1 \uvec{v} \bigr) \; , \\
\mathbf{c} & \coloneq \uvec{w} \; ,
\end{split}
\label{eq:rotatedbasisvectors}
\end{equation}
where $\mu_{\mathbf{k}} \!=\! \bigl( (2q_0L_1)^2 \!+\! (\varepsilon_f E \Delta_{\mathbf{k}})^2 \bigr)^{1/2}$ is a normalisation factor. In this rotated local frame, the choice of basis tensors
\begin{equation}
\begin{split}
\mathbf{M}_{\pm 2} & = \tfrac{1}{2} \bigl( \mathbf{b} \pm i \mathbf{c} \bigr) \otimes \bigl( \mathbf{b} \pm i \mathbf{c} \bigr) \; , \\
\mathbf{M}_{\pm 1} & = \tfrac{1}{2} \Bigl[ \mathbf{a} \otimes \bigl( \mathbf{b} \pm i \mathbf{c} \bigr) + \bigl( \mathbf{b} \pm i \mathbf{c} \bigr) \otimes \mathbf{a} \Bigr] \; , \\
\mathbf{M}_0 & = \tfrac{1}{\sqrt{6}} \bigl( 3\mathbf{a} \otimes \mathbf{a} - \mathbf{I} \bigr) \; ,
\end{split}
\end{equation}
results in the desired diagonalisation of $F^{(2)}$,
\begin{equation}
F^{(2)} =  \sum_{\mathbf{k}}  N_k^{-1} \sum_{m=-2}^2 Q_m^2(k) \Bigl\{ \tfrac{\tau}{4} + \tfrac{L_1}{2} k^2 - \tfrac{m}{2} \mu_{\mathbf{k}} k \Bigr\} \; .
\label{eq:f2}
\end{equation}

We see from Eqs. \eqref{eq:rotatedbasisvectors} that there is a
strong sense in which the transition from purely chiral to
purely flexoelectric coupling may be viewed geometrically as a $\pi /
2$ rotation. Indeed, this is already evident in the one dimensional
cholesteric and splay--bend states of Fig.~\ref{fig:distortions},
which are related by a $\pi /2$ rotation about the vertical direction. 
This rotation is none other than the well known flexoelectro-optic effect \cite{patel}. 

Given this formal similarity it is natural to ask whether a precise
correspondence can be made between the stable phases, and their
properties, in the purely chiral and purely flexoelectric limits. To this end we now focus our attention on the purely flexoelectric sector and analyse a number of structures motivated by the analogy with cholesterics. 

The simplest flexoelectric structure is a one dimensional phase that was described in Meyer's original paper \cite{meyer}, which we shall call splay--bend, see Fig.~\ref{fig:distortions}(b). The $\mathbf{Q}$-tensor appropriate to splay--bend is the analogue of that for the cholesteric helix in chiral liquid crystals:
\begin{equation}
\mathbf{Q} = \frac{Q_2}{\sqrt{2}} \begin{pmatrix} \cos (kx) & 0 & \sin (kx) \\ 0 & 0 & 0 \\ \sin (kx) & 0 & -\cos (kx) \end{pmatrix} - \frac{Q_h}{\sqrt{6}} \begin{pmatrix} -1 & 0 & 0 \\ 0 & 2 & 0 \\ 0 & 0 & -1 \end{pmatrix}
\end{equation}
where we have used the electric field to define the $z$-axis of a Cartesian coordinate system and taken the wavevector to be orthogonal to the field so as to maximise $\Delta_{\mathbf{k}}$ and hence minimise the free energy. The free energy is readily shown to be 
\begin{equation}
\begin{split}
F & = \tfrac{\tau}{4} \bigl( Q_2^2 + Q_h^2 \bigr) - 3Q_2^2Q_h + Q_h^3 + \bigl( Q_2^2 + Q_h^2 \bigr)^2 \\
& \quad + \tfrac{Q_2^2}{2} \bigl( L_1 k^2 - 2\varepsilon_f Ek \bigr) \; .
\end{split}
\label{eq:fesplaybend}
\end{equation}
It is formally identical to that of the cholesteric phase with
$\varepsilon_f E/L_1$ playing the role of the chiral parameter,
$2q_0$. Therefore the solutions for $Q_2$ and $Q_h$ as a function of the reduced temperature, $\tau$, and field strength, $E$, obtained by minimising Eq. \eqref{eq:fesplaybend}, are the same as those obtained for the cholesteric phase \cite{gareth}, provided one substitutes $2q_0$ with $\varepsilon_f E/L_1$.

\begin{figure}[tb]
\begin{center}
\includegraphics[width=82mm]{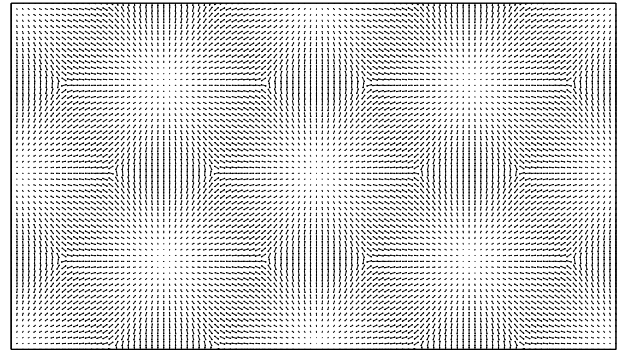}
\caption{Director field of the hexagonal flexoelectric blue phase described in the main text obtained from a numerical minimisation at paramter values $\kappa_{\text{E}} \!=\! 1, \tau \! -\! \kappa_{\text{E}}^2\! =\! 0$ where the phase is stable, see Fig.~\ref{fig:flexopd}. The structure is periodic in both directions.}
\label{fig:hexplanar}
\end{center}
\end{figure}

Extending the analogy with cholesterics, we now consider flexoelectric phases with a higher dimensional structure; flexoelectric analogues of the cholesteric blue phases. Based on the heuristic observations that the field picks out a preferred direction in space, and that the flexoelectric coupling vanishes for wavevectors parallel to the field, it seems likely that the free energy will be minimised by two dimensional structures, containing only wavevectors orthogonal to the field. Therefore we consider first a two dimensional phase possessing hexagonal symmetry which may be obtained by choosing the fundamental set of Fourier modes to be along the directions $\{ \pm \mathbf{e}_x, \pm (- \mathbf{e}_x + \sqrt{3} \mathbf{e}_y)/2, \pm (- \mathbf{e}_x - \sqrt{3} \mathbf{e}_y)/2 \}$. An approximate $\mathbf{Q}$-tensor, comprising only this fundamental set, all with $m=2$, and a homogeneous component, is given by
\begin{equation}
\begin{split}
& \mathbf{Q} = \frac{Q_2}{\sqrt{6}} \biggl\{ \tfrac{1}{2} \Bigl( \mathbf{e}_x + i \mathbf{e}_z \Bigr) \otimes^S  \text{e}^{-i(kx-\psi_1)} \\ 
& + \tfrac{1}{2} \Bigl( \tfrac{-1}{2} \mathbf{e}_x + \tfrac{\sqrt{3}}{2} \mathbf{e}_y + i \mathbf{e}_z \Bigr) \otimes^S  \text{e}^{-i(k(-x+\sqrt{3}y)/2-\psi_2)} \\ 
& + \tfrac{1}{2} \Bigl( \tfrac{-1}{2} \mathbf{e}_x - \tfrac{\sqrt{3}}{2} \mathbf{e}_y + i \mathbf{e}_z \Bigr) \otimes^S  \text{e}^{-i(k(-x-\sqrt{3}y)/2-\psi_3)} \\ 
& + \; \text{H.c.} \biggr\} + \frac{Q_h\text{e}^{i\delta}}{\sqrt{6}}\, \Bigl( 3\mathbf{e}_z \otimes \mathbf{e}_z - \mathbf{I} \Bigr) 
\end{split}
\end{equation}
where H.c. stands for Hermitian conjugate, $\psi_1, \psi_2, \psi_3 \! \in \! [0,2\pi)$,  $\delta \! \in \! \{ 0,\pi \}$ and the short-hand notation $\otimes^S$ denotes a symmetrised tensor product, i.e., $\mathbf{u} \otimes^S\! \coloneq \!\mathbf{u} \otimes \mathbf{u}$. The director field of this hexagonal flexoelectric blue phase is shown in Fig.~\ref{fig:hexplanar}. The structure consists of an hexagonal lattice of strength $-1/2$ disclinations separating regions in which the distortion is one of pure splay along any straight line passing through the centre of the hexagon. Indeed, expanding in cylindrical polars $(\rho,\phi,z)$, the director field in a local neighbourhood of the centre of each hexagon is
\begin{equation}
{\bf n} = \cos \Bigl( \tfrac{Q_2k\rho}{3Q_2 + 2Q_h} \Bigr) {\bf e}_z - \sin \Bigl( \tfrac{Q_2k\rho}{3Q_2 + 2Q_h} \Bigr) {\bf e}_{\rho} \; ,
\end{equation}
and therefore the structure may be aptly referred to as a {\em double splay cylinder} since it is the clear analogue of double twist cylinders in the cholesteric blue phases.

The free energy of the hexagonal flexoelectric blue phase is
\begin{equation}
\begin{split}
& F = \tfrac{\tau}{4} \Bigl( Q_2^2 + Q_h^2 \Bigr) - \tfrac{1}{4} \tfrac{2\varepsilon_f^2 E^2}{L_1} Q_2^2 - \tfrac{3 \text{e}^{i\delta}}{2} Q_2^2Q_h \\
& - \text{e}^{i\delta} Q_h^3 + \tfrac{27}{32} \cos (\psi_1 + \psi_2 + \psi_3) Q_2^3 + \tfrac{233}{192} Q_2^4 \\
& + \tfrac{7}{2} Q_2^2Q_h^2 + Q_h^4 - \tfrac{9}{8} \cos (\psi_1 + \psi_2 + \psi_3 + \delta) Q_2^3Q_h \; .
\end{split}
\label{eq:fehex}
\end{equation}
This is minimised by choosing the phases to satisfy $\psi_1 \!+\! \psi_2 \!+\! \psi_3 \!=\! \pi$ and $\delta \!=\! 0$. As in the case of the splay--bend phase, this is identical to the free energy of the cholesteric analogue, the planar hexagonal blue phase \cite{brazovskii,ghsa}. Consequently, we may draw on this previous work for cholesterics to conclude that the hexagonal flexoelectric phase has lower free energy than the one dimensional splay--bend phase at sufficiently large field strength. However, in the chiral case the hexagonal blue phase is not stable since one of the cubic blue phases always has lower free energy. It is therefore necessary to see whether a similar result also holds in the flexoelectric case.

Although one may reasonably expect that the structure of the cubic blue phases provides a good starting point for a consideration of flexoelectric blue phases, the latter differ in a number of respects. In particular, since the flexoelectric coupling is vectorial, it defines a preferred direction in space which will act to lower the symmetry from cubic to tetragonal. We therefore considered the set of possible structures obtained by selecting the Fourier modes of the ${\bf Q}$-tensor to be
\begin{align}
\Bigl\{ & {\bf 0}, \pm \tfrac{2\pi}{L_x} {\bf e}_x, \pm \tfrac{2\pi}{L_x} {\bf e}_y, \pm \tfrac{2\pi}{L_x} \bigl( {\bf e}_x + {\bf e}_y \bigr), \pm \tfrac{2\pi}{L_x} \bigl( -{\bf e}_x + {\bf e}_y \bigr), \notag \\
& \pm \tfrac{2\pi}{L_z} {\bf e}_z, \pm \tfrac{2\pi}{L_x} \bigl( {\bf e}_x + \tfrac{L_x}{L_z} {\bf e}_z \bigr), \pm \tfrac{2\pi}{L_x} \bigl( {\bf e}_y + \tfrac{L_x}{L_z} {\bf e}_z \bigr), \notag \\
& \pm \tfrac{2\pi}{L_x} \bigl( -{\bf e}_x + \tfrac{L_x}{L_z} {\bf e}_z \bigr), \pm \tfrac{2\pi}{L_x} \bigl( -{\bf e}_y + \tfrac{L_x}{L_z} {\bf e}_z \bigr) \Bigr\} \; , 
\label{eq:tetragonal}
\end{align}
where $L_x,L_z$ are the lattice parameters along the $x-$ and $z-$
axes of the conventional unit cell respectively. This set of
structures allows for considerable scope, containing as special cases
${\bf Q}$-tensors that are exact analogues of those representing the
cholesteric cubic blue phases as well as that of the two dimensional
square structure shown in Fig.~\ref{fig:square}(b). We have
investigated the free energy minima of these structures both using
approximate analytic calculations and with an exact numerical
minimisation of the free energy for a range of initial conditions, varying the ratio of the lattice parameters $L_x/L_z$ and the relative amplitudes and phases of the Fourier components. The numerical minimisation was performed using a lattice Boltzmann algorithm with an identical technique to that described in \cite{gareth}. In all instances a single minimum was obtained which corresponds to the two dimensional square flexoelectric blue phase of Fig.~\ref{fig:square}(b).  

\begin{figure}[tb]
\begin{center}
\includegraphics[width=60mm]{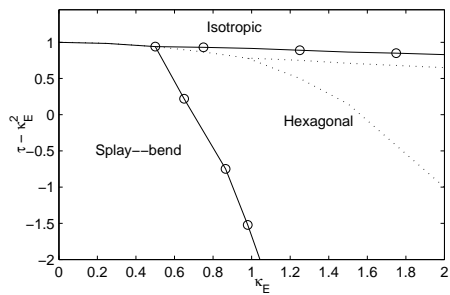}
\caption{Phase diagram of the flexoelectric blue phases. The circles and solid lines show the numerical phase boundaries, while the analytic results are given by the dashed lines.}
\label{fig:flexopd}
\end{center}
\end{figure}

These results were combined with the analytic expressions for the free energy of the splay--bend and hexagonal structures, Eqs.~\eqref{eq:fesplaybend} and \eqref{eq:fehex}, to construct an approximate phase diagram. We find that although the two dimensional square phase is a local minimum of the free energy within the set of structures defined by \eqref{eq:tetragonal}, it never has lower free energy than the hexagonal phase. The same result is obtained numerically, leading to the phase diagram shown in Fig.~\ref{fig:flexopd}, which has been plotted in the $\kappa_{\text{E}}, (\tau \!-\! \kappa^2_{\text{E}})$ plane, where $\kappa_{\text{E}}\! \coloneq \!\sqrt{2 \varepsilon_f^2 E^2 / L_1}$, in accordance with the conventions adopted in cholesterics \cite{gareth}. As the electric field strength is increased the hexagonal flexoelectric blue phase becomes stabilised over an increasing temperature interval between the isotropic and splay--bend phases. The discrepancy between the numerical and analytic phase boundaries is similar to that found in cholesterics and arises from neglecting higher order wavevector harmonics in analytic calculations. In the cholesteric blue phases the lattice periodicity is not identical to the cholesteric pitch, but is generally somewhat larger. From the numerical minimisation we find the same feature in the hexagonal flexoelectric blue phase, with the lattice parameter being approximately $15 \%$ larger than the `pitch' ($\pi L_1 / \varepsilon_f E$) of the splay--bend phase at the triple point.

The phase diagram bears a qualitative resemblance to that of cholesterics \cite{gareth}, but the details reveal differences between chiral and flexoelectric couplings. In particular, it seems that the suppression of wavevectors parallel to the field introduced by the quantity $\Delta_{{\bf k}}$ is sufficient to convert the fully three dimensional states which are stable at large chiral coupling into two dimensional states under flexoelectric coupling.

We comment that the local director field in the double splay regions of the flexoelectric blue phases is the same as that in the escape configuration lattice phase recently suggested by Chakrabarti {\it et. al.} \cite{chakrabarti}. However, the mechanism by which the structure occurs, and is stabilised, is different. Their proposed structure arises without the application of an electric field, being instead stabilised by the saddle splay elastic constant and including weak anchoring of the director field and surface tension at the interface between the nematic and isotropic regions of the fluid. Note also that the hexagonal symmetry of the phase we obtain is different to the square symmetry predicted in that work.

It is important for potential experimental work to give an estimate of
the field strength required to stabilise the two dimensional hexagonal
flexoelectric blue phase. A good estimate is that the strength of the flexoelectric coupling, $\varepsilon_f E$, should be as large as the chiral coupling, $2q_0L_1$, in a compound which possesses cholesteric blue phases. This gives a field strength $E \approx 2q_0L_1/ \varepsilon_f \approx \pi K/ep$, where we have replaced the Landau--de Gennes parameters with their director field equivalents; an average Frank elastic constant, $K$, and the average of the flexoelectric coupling constants, $e\! \coloneq \!(e_s + e_b)/2$. $p$ is the pitch in the cholesteric phase of a compound which also displays blue phases, typically $\sim 0.3 \mu m$. Recently materials have been developed for flexoelectric purposes and found to have a large flexoelastic ratio $e/K\! \sim \!1 V^{-1}$ \cite{coles2}. Thus in these materials the required field strength is expected to be $E\! \sim \!10 V \mu m^{-1}$, which is within the experimentally accessible range. In addition, for the proposed flexoelectric blue phases to be seen, it will be helpful to use a material with as close to zero dielectric anisotropy as possible and to take care to ensure that surface effects are negligible compared to the bulk. The analogy with cholesterics further suggests that flexoelectric blue phases are only likely to be found in a very narrow temperature range ($\sim \! 1$ K) just below the isotropic--nematic transition temperature, although this is predicted to expand with increasing field strength. 

\vspace{0.1mm}
We thank Davide Marenduzzo for useful discussions.

\end{document}